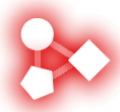

# Designing a Web-Based Interactive Audio Library Automation System for Visually-Impaired People and Evaluation of Its Usability

**Aslihan Tufekci**
*Gazi University,*
*Gazi Faculty of Education,*
*Dept. of Computer Instruction Technologies, Turkey*
*asli@gazi.edu.tr*

**Yahya Balaman**
*Ankara Abidinpasa Technical and Industrial High School, Turkey*
*yahyabalaman@gmail.com*

**Utku Kose**
*Usak University,*
*Computer Sciences App. & Res. Center, Turkey*
*utku.kose@usak.edu.tr*

## Abstract

The aim of this study is to introduce an application that enables information sharing and communication between visually-impaired individuals and able-bodied. For the purposes of the study, web-based audio library automation was designed and the usability of the system was analyzed regarding the volunteers who record audio books and the visually-impaired individuals. The visually-impaired individuals who took part in the test procedures in order to make a general evaluation of the system reported that the system was theoretically necessary and successful. As for the usability aspect, positive comments were received regarding the automation system developed. The authors believe that the current study is likely to be an alternative reference source for the related literature and further research studies to be conducted in the field.

## 1. INTRODUCTION

What lie behind most of the innovations and developments shaping and guiding our daily lives is rapid and influential changes in the field of technology. Such changes, which might be analyzed in terms of technological advancements and improvements, have brought many advantages in terms of accessing and making use of information. In this respect, especially the significant developments and improvements in computer technology have altered today's life standards radically. When the current conditions are considered, it is clearly seen that accessing information is relatively easier, more effective and faster compared to the past due to increasing use of tablet PCs, smart Phones and laptops as well as other technological advances enabling satisfactorily high speed internet access. The traditional printed materials that were available in libraries in the past such as books, magazines and newspapers have been replaced by e-books. Nowadays, if you have access to the Internet, the information you want to get is only one click away within seconds thanks to e-book applications. Developed within the framework of technological advances, e-book applications are not, in fact, convenient for the handicapped while they facilitate the lives of able-bodied in terms of accessing and using information. To illustrate with, although the visually-impaired are able to listen to text-based e-books thanks to screen-readers, this application is not possible for illustration-based e-books.

From a broader point of view, it is clear that there is demand for certain regulations and improvements targeting the visually impaired individuals when accessing and using information is considered; especially regarding computer-based applications used in libraries. This demand is voiced in legal documents as well. For instance, Article 5th of "the Declaration on the Rights of Disabled Persons" published on December 9th 1975 reads: "The handicapped individuals have the right to demand that necessary precautions should be taken that may help them to become a self-sufficient individuals" (United Nations, 1975). As highlighted in this article, the disabled have the right to receive education like all other individuals and to demand that education materials should be adapted to meet their needs". Similarly, the fact that visually-impaired individuals need different regulations and improvements is also reflected in the definitions given in the "Regulation for Private Education Institutions" published by Turkish Ministry of Education. The following definitions write: "Blindness is defined as visual acuity of less than 20/400 (6/120), or corresponding visual field loss to less than 10 degrees, in the better eye with best possible correction and therefore such people cannot benefit from visual elements of education process provided in educational institutions" and "low vision means having the visual acuity between 10 and 30 degrees in the better eye after the best possible correction and therefore not being able to benefit from the visual elements of education process" (Turkey Ministry of National Education, 2008)

Several institutions and foundations, which are volunteers to use technological advances to the advantage of the visually-impaired, have already conducted some studies and launched projects on the topic. In parallel with the concept dealt with in this current study, some special sections were designed for the visually-impaired in many libraries in Turkey and the world. The recorded versions of considerable number of printed materials are, now, available in these special sections prepared for the visually impaired in various forms such as tape cassettes and CD-ROMs etc. These recordings are done by volunteers in professionally-equipped studio environments. Unfortunately, the number of audio books has not reached a desired level yet due to certain limitations such as the insufficient number of volunteers, inconvenient transportation to the studios, and inadequate number of available studios. As an alternative solution to this problem, some associations and foundations provided opportunities for the volunteers to do the recordings by installing a microphone and sound-recording software into their own PCs and later to upload these files to the internet so that the visually-impaired were able to listen to these recordings. The problem with this application was that the quality of the recordings was far lower than professional studio recordings. This method produced more audio books than professional studios. However; in practice, many problems were encountered. For instance, the volunteers assumed that they recorded the audio books into CD-ROMs and DVD-ROMs, but the CDs and DVDs sent were blank unfortunately. Similarly, some CDs and DVDs were damaged during shipping, mailing costs were high and there was not enough number of staff to listen to the recordings and approve them for use.

The aim of this study is to present an application approach that enables information sharing and communication between visually-impaired individuals and able-bodied in a way to produce solutions for the problems mentioned above. In this regard, web-based audio library automation was designed and the usability of the system was analyzed in terms of volunteers doing the recordings and visually-impaired individuals. The automation system mentioned above enables the volunteer members to transfer the audio books recorded into the server via FTP. At this point, uploaded files are listened to online by authorized volunteers and later approved or rejected. The visually impaired individuals can listen to approved recordings either online or by downloading them into their PCs, therefore; not only a group of visually-impaired individuals living in a specific location but also all the visually-impaired having internet access can benefit from this service. All the functions and uses mentioned above are efficiently available thanks to user-friendly, fast and effective interfaces. In this regard, the system designed and developed is likely to contribute considerably to the related literature and further research studies to be conducted in the field.

## 2. AUDIO LIBRARY CONCEPT AND APPROACH

The features and functions of the automation system designed and developed within the framework of this current study are inspired by the concept and approach called "Audio

Library", as it is called in the related literature. At this point, it would be useful to give some basic information about "Audio Library" concept and approach so that the readers will have a better understanding of the study and the concept.

## 2.1 Audio Book

When a book is read aloud and recorded on a media, this new version is called "audio book" or "talking book". "Audio books" are not only for the visually-impaired but also those who are not able to read printed books due to their heavy workload and long hours of commuting. In many European countries and the USA, professionally recorded versions of printed books are sold, and CD-ROMs are preferred the most by the book worms in order to listen to the books especially while driving their cars.

Audio books for the visually-impaired are mostly read aloud and recorded by volunteers. This reading aloud is digitally recorded by using computer technology and generally in sound recording rooms established specifically for this purpose. In this regard, there are sound recording rooms in universities and "the sections for the blind" of public libraries and also in some institutions founded specifically for that purpose. Alternatively, any volunteers can record "audio book" at his / her home by using computer software and a microphone very easily (Corbeil & Valdes-Corbeil, 2007).

## 2.2 Audio Library

Physical or virtual environments established to display "audio books" are called "Audio Library". Today, many libraries set up special units where the handicapped individuals have the opportunity to listen to audio books by using some technological devices such as computer, type recorder etc. Since it is not convenient for the visually-impaired to visit libraries, they can access these audio book units of libraries from their homes via the internet. In addition to state-run public libraries, some foundations and private institutions also provide "audio book" services via the internet. At this point, due to copyrights issue, it has been necessary to make certain amendments in the related regulations in order to facilitate the access by the visually-impaired who would like to benefit from these audio books. Another important issue in this respect is that the visually-impaired individuals who would like to receive this service are required to submit their identification cards or doctors' report proving their visual-impairment to the authorities working for these libraries to ensure a problem –free access (Turkey-Official Gazette, 1951).

## 2.3 Audio Book Applications

A detailed examination of the related literature shows that there are a lot of institutions and organizations that provide audio library service in various ways. Several examples of these organizations and institutions and some details about the practical applications of these services are as follows:

- The USA-based "National Library Service for the Blind and Physically Handicapped (NLS)" provides "audio books" in cartridges. The visually-impaired individuals can receive the audio book they want to read in a cartridge via snail-mail if they provide the details about their addresses. It is necessary to have a special "sound-reader" device to which the cartridges that include the audio book(s) must be inserted in order to listen to these audio books (The Library of Congress, 2012).

- "Recording for the Blind and Dyslexic" broadcast from a private university located in New Jersey, USA. The website provides services for both able-bodied and the handicapped. The audio books can be searched with the help of a search engine available in the website and purchase order can be given. The website charges money for all the audio books and requires membership to make the purchase (Recording for the Blind and Dyslexic, 2012).

- "San Francisco Public Library", USA, has a special unit for the visually-impaired. Thanks to a device available in this unit, the books written in Braille Alphabet can be recorded effectively. Another device enables the written words on the screen to be read aloud for the visually-impaired. Similarly, the parts of the books can be printed out in Brille Alphabet format. The other services provided in the library are as follows: the books are projected on a TV screen in very large fonts for those who has very weak sight; and finally, a device called VERA (Very Easy Reading Appliance) is used to scan the books as texts and later the scanned texts are recorded on various media tools (San Francisco Public Library, 2012).

- "Maryland State Library for the Blind and Physically Handicapped", based in the USA, houses the audiotapes and CD-ROMs specifically prepared for the visually-impaired. Since the library cannot provide any services via the internet, those who would like to access these materials have to go to the library and use these special units organized for them. The recorded copies of many written texts such as stories, novels, poem books etc are available for the users on demand (Maryland State Library for the Blind and Phisically Handicapped, 2012).

▪ "Royal National Institute of Blinds", located in England, has 12.000 books for the visually-impaired in its inventory. Called "Talking Book Service", this service also includes the translated versions of these books in many languages. The books are sent via snail-mail to the visually-impaired individuals who cannot come to the library for apparent reasons. This audio library service is one of the many services provided to make the visually-impaired individuals' lives easier. It is reported that more than half of the libraries in England has units providing services for the visually-impaired (Bakırcı, 2009).

▪ In the Turkish context, there are a lot of institutions and organizations providing audio library service. Among these institutions, organizations and some related applications are "Talking Book Portal in National Library – The Center for the Visually-impaired", "Audio Book Libraries - Altınokta Körler Derneği" (Altınokta Association for the Blind), "Audio Library Portal – İstanbul Metropolitan Municipality", "Audio Book Volunteers Website – Boğaziçi University" and "Turkish Library for the Visually-impaired". (Istanbul Metropolitan Municipality, 2012; SKG – Sesli Kitap Gönüllüleri – Audio Book Volunteers, 2012; TÜRGÖK – Türkiye Görme Özürlüler Kitaplığı - Turkey Visual-impaired People Library, 2012).

A closer look into the issue shows that some of the audio library services are provided only in the units specifically designed for the visually impaired in the buildings of the foundations and libraries. Therefore; this situation causes some problems for the visually-impaired that do not reside in the cities having foundations and libraries or live far from these buildings. Similarly, since the recordings are done in the studios available in these institutions, volunteers also find it difficult to come to these studios. However, this problem seems to have been solved thanks to online audio books services provided via websites by these institutions.

## 3. AUDIO LIBRARY AUTOMATION SYSTEM

This section presents information about the basic features and functions of audio library automation system designed for the purposes of enabling the visually-impaired to access the books recorded by volunteers in online environments. Within the general framework of the system, various mechanisms such as interfaces, programs and web services have been developed to enable volunteers to upload the recorded audio book files to the website's database and to use all other functions of the system. At this point, the features and functions will be explained based on the interfaces forming the overall system and automation system.

### 3.1 Audio Library Website

Audio Library Website is an interactive platform in which volunteers are able to upload to the server the recordings they complete at their homes by using a computer and the administrators are authorized to approve or reject these recordings by listening to them one by one. Later, this platform enables the visually-impaired to listen to the recordings online or by downloading them into their PCs. The following is the detailed information about the interfaces used in this platform:

#### 3.1.1 User log-in interface

"User Log-in Interface" refers to password-operated log-in screen that is used by the handicapped, volunteers and administrators to access the website (Figure 1).

| Gönüllü Üye Girişi | Engelli Üye Girişi | Yönetim Giriş |
|---|---|---|
| Kullanıcı Adı :    Şifre :    Giriş | Kullanıcı Adı :    Şifre :    Giriş | Kullanıcı Adı :    Şifre :    Giriş |
| | Şifremi Unuttum     Yeni Kayıt | |

**Figure 1.** User Log-in Interfaces

The interface also includes the areas where membership applications can also be made by the handicapped and volunteers. As for the application process for the volunteers, they are required to submit one-minute trial sound-recording in addition to the form to be filled out. Filling out the form is enough for the handicapped to be a member of the website.

#### 3.1.2 Home page interface

Home page interface displays information regarding the function of the website and other necessary details to persuade the visitors to be volunteers. In addition, recent news and announcements addressing the handicapped are published on the home page. The visitors can write their ideas and comments on "visitors' page" or can read what the previous visitors have written as well (Figure 2).

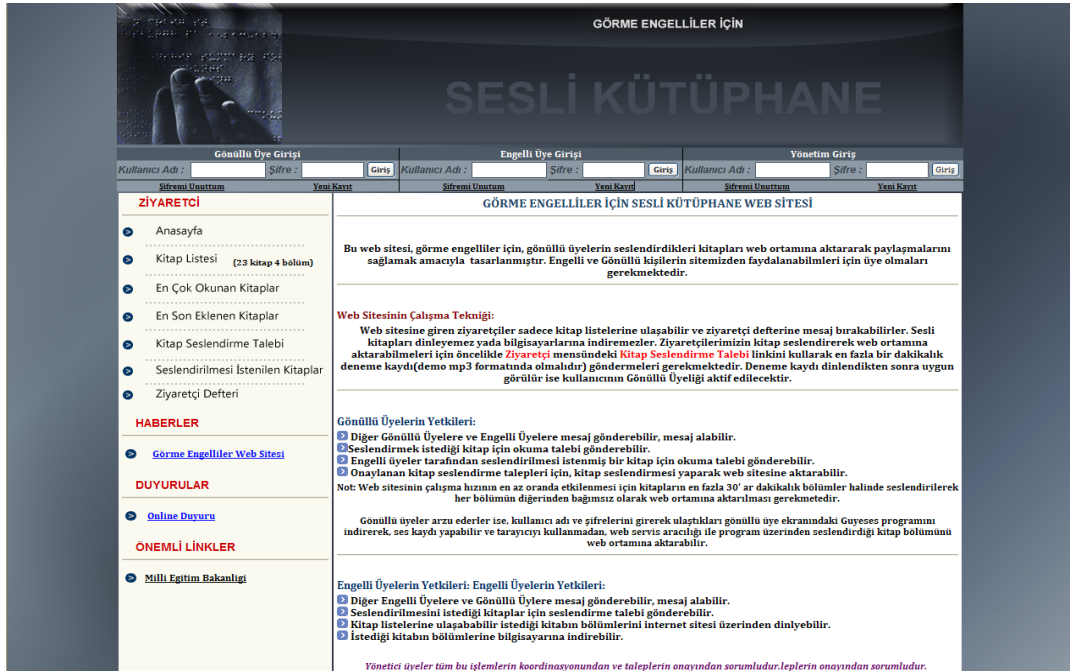

**Figure 2.** Home Page Interface

### 3.1.3 Volunteer Member Interface

The page displayed after the volunteer members log-in by entering their user names and passwords is called "volunteer member interface". (Figure 3)

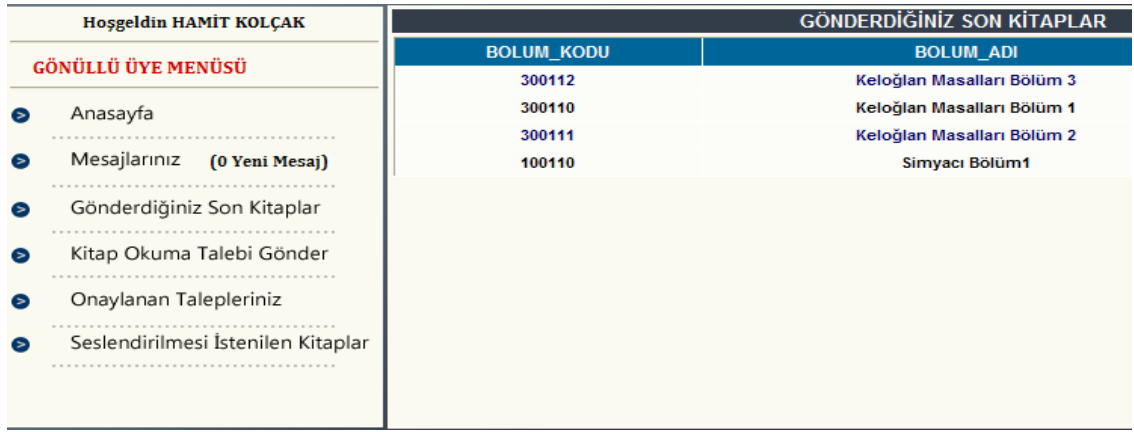

**Figure 3.** Volunteer Member Interface

Volunteer members are able to manage their in-site messages in this area. The presence of two separate recordings of the same book on the website is not allowed since it brings extra technical load to the site and the volunteers are more easily directed to the books that have not been recorded yet. Any volunteer member who would like to record a book should inform the website's admins about this demand and later choose one of the books in the list provided by the handicapped members regarding the books to be recorded. The next step is that admins evaluate this demand for recording and approve it if they don't see any problems. After the approval, the page to which the volunteer can upload the recorded files are activated in the volunteer's page (Figure 4).

| SEÇTİĞİNİZ KİTAP İÇİN DAHA ÖNCE GÖNDERDİĞİNİZ ONAYLANMIŞ BÖLÜMLER | | | | | |
|---|---|---|---|---|---|
| | **K_KOD** | **BOLUM_KODU** | **BOLUM_ADI** | **SURE** | **EKLENMETARIHI** |
| SEÇ | 3001 | 300110 | Keloğlan Masalları Bölüm 1 | | 20.04.2012 09:00:00 |
| SEÇ | 3001 | 300111 | Keloğlan Masalları Bölüm 2 | | 20.04.2012 09:00:00 |

Toplam 2 Adet Gönderilmiş Bölüm Bulundu.

| SEÇTİĞİNİZ KİTAP İÇİN GÖNDERDİĞİNİZ ONAY BEKLEYEN BÖLÜMLER | | | | | |
|---|---|---|---|---|---|
| | **K_KOD** | **BOLUM_KODU** | **BOLUM_ADI** | **SURE** | **EKLENMETARIHI** |
| SEÇ | 3001 | 300112 | Keloğlan Masalları Bölüm 3 | | 20.09.2008 09:00:00 |

**BÖLÜM AYRINTILARI**

| | | | |
|---|---|---|---|
| *Kitap Kodu :* | 3001 | *Bölüm Kodu :* | 300113 |
| *Bölüm Adı :* | Keloğlan Masalları Bölüm 3 | *Süre :* | |
| *Eklenme Tarihi :* | 08.09.2009 | *Gönderen :* | HAMİT KOLÇAK |
| *Bölüm Dosya :* | | Gözat... | |

gönder

**Figure 4.** Volunteer Member Interface

Volunteers have the options to send the recording as one single file or in parts. By doing so, slow downloads and bufferings due to large files can be avoided.

### 3.1.4 Interface for the handicapped

"Interface for the Handicapped" is the page that is displayed after the handicapped individuals access by using their user names and passwords. They are also able to manage their accounts and in-site messages (Figure 5).

**Figure 5.** The interface for the Handicapped - Recordings Demands

In addition, the handicapped members are able to send their demands for the books they want to be recorded by adding the names of these books to the list called "the books demanded by the handicapped to be recorded". This list is displayed on the volunteers' page. By using their own accounts, the handicapped members can access all audio book contents recently uploaded to the website and listen to them online or download them into their PCs.

**Figure 6.** The interface for the Handicapped – Section for "Listening to Audio Book"

### 3.1.5 Admin interface

Admin interface is defined as the platform designed for the admins responsible for the management of the website. They are mainly responsible for managing the membership procedures and audio book transfer traffic. By using admin interface, it is possible to examine the forms filled out by the handicapped and volunteers and later to approve their memberships and finally to send the usernames and the passwords to their e-mail addresses.

The most important function of admins is to manage audio book traffic. After volunteers submit their wish to record books to the admins, the admins check whether or not this book has been recorded before or the sound quality is sufficient. Volunteer members can start recording only after the approval by the admins. The parts of the book recorded can be uploaded to the website by using the user-specific pages or software that can be downloaded from the site.

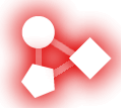

**Figure 7.** Admin Interface - Home Page

Each recording received is listened by the admins and later approved or rejected. Moreover, admins are authorized to publish news, announcement and links as well as to delete or correct them.

## 3.2 Volunteer Member Software

"Volunteer Member Software" is an application designed to enable volunteer members to follow the procedures without using a web browser. Thanks to this program, volunteer members can carry out each action by using a web interface after they log in by entering their user names and passwords. With the addition of mp3 recording software, volunteer members now don't have to search for similar software to record audio books.

Volunteer member software can be downloaded by the users who log in to the website as a volunteer member. The software can easily be installed by the users simply by following the installations steps after the download.

### 3.2.1. The features of the software and functions

When volunteer member software is run, log-in panel is displayed first. In order to access the system, volunteer members are required to log in by entering the same username and the password they use to enter the website (Figure 8).

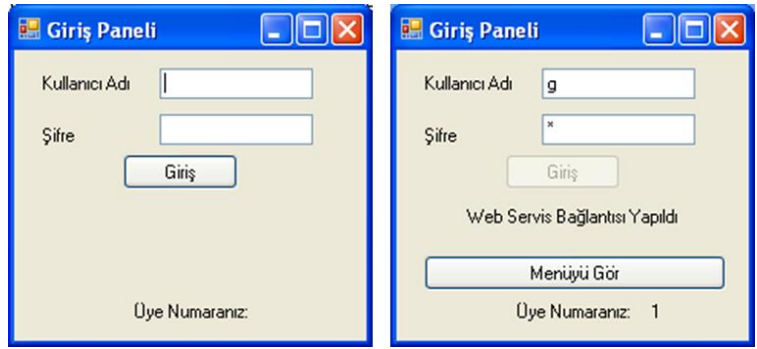

**Figure 8.** Volunteer Member Program and Sign Up Panel

Volunteer member software can access the database and sound files available in the website by using a kind of Web service approach. Therefore, volunteer members can access all the information in their own user-specific pages in the website through "Recorded File Submission Screen" in the software and can submit the parts of the audio books they recorded accordingly (Figure 9). Except from currently employed approach, spreading of the recorded files is thought to be supported with also Web 2.0 based podcasting (Deans, 2008; Köse, 2010) in a short time.

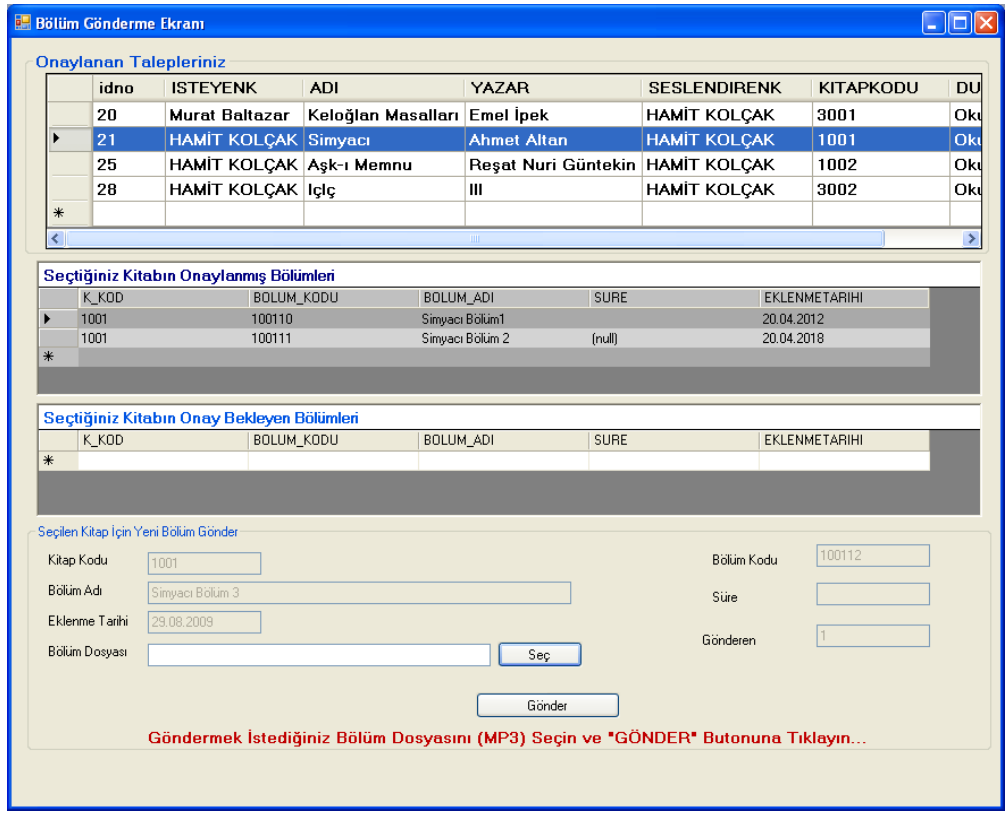

**Figure 9.** Volunteer Member Software - Recorded Data Submission Page

### 3.3. The Need Analysis for the Visually Impaired and the Findings

Prior to the design process of automation system, a survey was administered to obtain data about the expectations of the visually-impaired from an audio library website study. It is clear that the data obtained from the survey study have been used for the design and development processes so that the system would meet the demands effectively. The items of the survey, the data obtained from each item and the related comments are explained in the following sections.

**Table 1.** Need Analysis Survey for the visually-impaired – items and the data obtained

| Survey Item | f | % |
|---|---|---|
| Convenient access to the website for the visually impaired. | 5 | 15 |
| Recording done at normal speed and correct in terms of punctuation rules. | 10 | 31 |
| User-friendly download option for audio books. | 2 | 6 |
| A great variety of audio books in the website. | 11 | 34 |
| Giving priority to the audio books demanded by the visually impaired. | 1 | 3 |
| Availability of audio book search option in the website. | 2 | 6 |
| The necessity that the ones who will do the recording should be knowledgeable about the process. | 3 | 9 |

According to the results displayed in Table 1, "a great variety of audio books" is the most popular expectation of the visually-impaired from the audio library website with a percentage of 34. This item has been the most important reason triggering the idea of the automation system. Since the number of audio books obtained through the traditional method is not sufficient, the ultimate aim of the system is to record as many books from different genres as possible in a short time by enabling volunteers to do the recording at their homes rather than in studios.

The second popular demand by the visually-impaired is "recording done at normal speed and correct in terms of punctuation rules" with a percentage of 31. As a function of the automation system admin interface, the recorded files submitted by volunteers are listened to and approved by the admins before they are serviced to the visually-impaired. The recordings that are rated as "insufficient quality" are rejected by the admins.

Considering the demand voiced by the visually-impaired, that is "convenient access to the website for the visually impaired", the designers used "ComboBox controls" for the interface designed for the visually-impaired since this is the easiest way for them to read the screen through screen-reader software.

As a result, it has been possible to design above mentioned automation system in a way to meet the demands and expectations of the visually-impaired effectively. It is also notable that using such system in mobile systems will provide many advantages. So, there are some works (not reported in this study) to adapt the system to a mobile interface similar to the one introduced previously (Tüfekçi *et al.*, 2013). For the currently introduced system, another issue to be considered in the process is "evaluation of the system usability".

## 4. EVALUATION

In order to test the efficiency and the effectiveness of the audio library automation system, various evaluation methods were used to follow a disciplined evaluation process. In this respect, usability tests and satisfaction surveys were included in the process respectively. The detailed explanations regarding evaluation approaches mentioned above will be given in the following sub-sections.

### 4.1 Usability Tests

Within the framework of usability test approach, different versions of usability tests were used for each user profile registered in the system. These usability tests were administered to a total of 15 participants, 3 for each profile category, since Nielsen states that 75 % of the usability problems can be identified by giving the usability test to 5 participants (Barnum, 2003).

Usability tests were given to the following participants according to the profile categories: 2 high school students attending the last year's courses and 3 teachers for "volunteer member interface"; 5 teachers for "admin interface"; and 5 visually-impaired individuals for "the interface for the handicapped". For the purposes of the tests, the participants were required to complete the tasks given as a list. Three different task lists were prepared for three different interfaces.

Before the application of usability tests, all the participants were provided with some basic information regarding the website's organization, its design and how they should behave during the administration of the tests. The researchers were careful with having as little interaction with the participants as possible during the administration of the tests. The tests were applied in a computer lab with internet access by using a computer equipped with a microphone and speaker. In order to increase test reliability and to do an easier analysis of usability tests results, the sessions were recorded by video cameras after taking the consent of the participants for recording. The run time started when participants started to read the question and finished when he / she successfully completed the task or passed to another tasks when she was convinced that she won't succeed. File uploading procedures and page streaming times due to slow internet access were not included in the run time.

### 4.2 The Data Obtained from Usability Tests

The data obtained from usability tests were used in order to evaluate the system in terms of efficiency and effectiveness. At this point, it is possible to categorize the data obtained from the tests as follows: the time spent to complete the task (seconds), task completion status (+ for the completed tasks and – for the incomplete ones). The tasks for each user profile and the data obtained for each task are presented in Table 2, Table 3 and Table 4 respectively.

**Table 2.** Volunteer Member Interface Usability Test Tasks and the Data Obtained

| Tasks | Time Spent (sec) | | | | | | Task Completion Status | | | | | |
|---|---|---|---|---|---|---|---|---|---|---|---|---|
| | A | B | C | D | E | Average time | A | B | C | D | E | % |
| Membership Application. | 137 | 106 | 108 | 147 | 116 | 123 | + | + | + | + | + | 100 |
| Member Log-in to the Website. | 95 | 87 | 119 | 106 | 143 | 110 | + | + | + | + | + | 100 |
| Changing user name and password. | 38 | 46 | 33 | 41 | 26 | 37 | + | + | + | + | + | 100 |
| Sending messages within the website. | 45 | 74 | 69 | 54 | 31 | 55 | + | + | + | + | + | 100 |
| Adding people to the Friend List. | 5 | 25 | 8 | 6 | 5 | 10 | + | + | + | + | + | 100 |
| Submitting the Demand for Recording a Book. | 33 | 21 | 24 | 33 | 22 | 27 | + | + | + | + | + | 100 |
| Reading the messages in the inbox. | 16 | 67 | 84 | 68 | 58 | 59 | + | + | + | + | + | 100 |
| Submitting the Recorded Files. | 48 | 38 | 68 | 61 | 44 | 52 | + | + | + | + | + | 100 |
| Submitting the Demand for the Books to be recorded. | 72 | 28 | 35 | 19 | 31 | 37 | - | + | + | + | + | 80 |
| Displaying the list of "the mostly read books". | 54 | 104 | 58 | 75 | 34 | 65 | + | + | + | + | + | 100 |
| Displaying the list of "the most recently added books". | 37 | 41 | 20 | 56 | 28 | 36 | + | + | + | + | + | 100 |
| Using Visitors' Page. | 72 | 69 | 72 | 59 | 64 | 67 | + | + | + | + | + | 100 |

As shown in Table 2, task completion success of the participants is 100 % except one task. Although there are differences among task completion times, the participants completed almost every given task. The differences in task completion times are thought to be due to the differences among the participants in terms of their internet surfing experiences, therefore they might have spent more time while trying to find the icons and links on the webpage.

**Table 3.** Admin Interface Usability Tasks and the Data Obtained

| Tasks | Time Spent (sec) | | | | | | Task Completion Status | | | | | |
|---|---|---|---|---|---|---|---|---|---|---|---|---|
| | A | B | C | D | E | Average time | A | B | C | D | E | % |
| Logging in to admins' panel. | 10 | 6 | 6 | 7 | 12 | 8 | + | + | + | + | + | 100 |
| Trial record procedures. | 100 | 65 | 67 | 79 | 64 | 75 | + | + | - | + | + | 80 |
| Approval of the sign-up of visually impaired member candidates. | 44 | 28 | 11 | 22 | 6 | 22 | + | + | + | + | + | 100 |
| Listening to audio boks online. | 74 | 30 | 32 | 80 | 50 | 53 | + | + | + | + | + | 100 |
| The procedure for the | 226 | 143 | 88 | 84 | 26 | 113 | + | + | + | + | + | 100 |
| demand to record a book. | | | | | | | | | | | | |
| The procedures for the books with completed recording. | 132 | 49 | 88 | 95 | 117 | 96 | + | + | + | + | + | 100 |
| The procedures for the approval of the recorded books. | 82 | 67 | 0 | 70 | 45 | 53 | + | + | - | + | + | 80 |
| Visitors' page operations. | 57 | 53 | 90 | 41 | 41 | 56 | + | + | + | + | + | 100 |
| Announcements operations. | 75 | 74 | 38 | 63 | 54 | 61 | + | + | + | + | + | 100 |
| News operations. | 114 | 65 | 85 | 75 | 57 | 79 | + | + | + | + | + | 100 |

When Table 3 is examined, it can be seen that all the participants successfully completed all the tasks; except "trial recording" and "Parts of books waiting for the approval" tasks. The differences among task completion times are due to differences in individual performances and the time wasted while downloading the audio books into the computer due to relatively slow internet speed.

**Table 4.** Usability Test Tasks (The Interface for the Visually Impaired and the Data Obtained)

| Tasks | Time Spent (sec) | | | | | | Task Completion Status | | | | | |
|---|---|---|---|---|---|---|---|---|---|---|---|---|
| | A | B | C | D | E | Average time | A | B | C | D | E | % |
| Submitting membership request. | 210 | 165 | 238 | 190 | 152 | 191 | + | + | + | + | + | 100 |
| Signing up to the interface for the visually impaired. | 80 | 190 | 95 | 65 | 52 | 96 | + | + | + | + | + | 100 |
| Changing user name and password. | 106 | 91 | 81 | 94 | 72 | 89 | + | + | + | + | + | 100 |
| Reading the messages. | 96 | 62 | 68 | 81 | 77 | 77 | + | + | + | + | + | 100 |
| Adding to friend list. | 120 | 84 | 79 | 65 | 80 | 86 | + | + | + | - | + | 80 |
| Listening to parts of books online. | 116 | 68 | 52 | 64 | 88 | 78 | + | + | + | + | + | 100 |
| Downloading parts of books into the computer. | 80 | 130 | 86 | 140 | 102 | 108 | - | + | - | + | + | 60 |
| Submitting a demand for a book to be recorded. | 148 | 84 | 45 | 68 | 82 | 85 | + | + | + | + | + | 100 |

As Table 4 clearly shows, the handicapped participants successfully completed all the tasks except two of them. The participants were observed to find it difficult to find the links required to complete the tasks since they were nervous while doing their tasks.

### 4.3 Satisfaction Surveys

After the administration of the usability tests, the participants were given different satisfaction surveys for each user profile. The aim of these surveys was to give the participants opportunity to express their experiences more effectively and to use the obtained data for the evaluation process. The surveys included 10 items for the "volunteer member interface" and "admin interface" but 7 items for "the interface for the handicapped". The participants filled out these surveys by expressing their opinions on a five-point Likert scale. The results are shown in Table 5, Table 6 and Table 7 and the total points given by the participants for each item are displayed in the column "Total" and the percentage value that can be taken according to the maximum point is displayed in the column "percentage".

**Table 5.** Volunteer Member Interface Satisfaction Survey Items and the Data Obtained

|   | Items | A | B | C | D | E | TOTAL | % |
|---|---|---|---|---|---|---|---|---|
| 1 | I was easily be able to manage the website functions. | 5 | 4 | 5 | 4 | 5 | 23 | 92 |
| 2 | It is easy to learn how to use website effectively. | 3 | 5 | 5 | 5 | 5 | 23 | 92 |
| 3 | "Help" messages are sufficient. | 4 | 4 | 4 | 4 | 4 | 20 | 80 |
| 4 | I was able to correct the mistakes very easily. | 5 | 4 | 5 | 4 | 4 | 22 | 88 |
| 5 | I am satisfied with the visual design. | 4 | 5 | 4 | 5 | 5 | 23 | 92 |
| 6 | The fonts used made the reading easier. | 4 | 3 | 4 | 4 | 4 | 19 | 72 |
| 7 | The colors used made reading easier. | 4 | 4 | 5 | 4 | 4 | 21 | 84 |
| 8 | I completed the tasks effectively. | 4 | 5 | 4 | 5 | 5 | 23 | 92 |
| 9 | The successive pages were easy to follow and understand. | 5 | 4 | 3 | 4 | 4 | 20 | 80 |
| 10 | I am satisfied with the website in terms of use. | 5 | 5 | 5 | 5 | 5 | 25 | 100 |

When Table 5 is examined, the volunteer members are found to have given less feedback compared to others for the item "The fonts used made the reading easier". Taking this feedback into consideration, the font used in the website was replaced by Trebuchet MS, since this font is more easily readable and more attractive. More than 80 % of the participants provided positive feedback for the other items.

**Table 6.** Administrator Interface Satisfaction Survey Items and the data Obtained

|  | Items | A | B | C | D | E | TOTAL | % |
|---|---|---|---|---|---|---|---|---|
| 1 | I was easily be able to manage the website functions. | 4 | 5 | 5 | 4 | 4 | 22 | 88 |
| 2 | It is easy to learn how to use website effectively. | 5 | 4 | 5 | 4 | 4 | 22 | 88 |
| 3 | "Help" messages are sufficient. | 5 | 5 | 5 | 4 | 4 | 23 | 92 |
| 4 | I was able to correct the mistakes very easily. | 3 | 4 | 4 | 5 | 5 | 21 | 84 |
| 5 | I am satisfied with the visual design. | 4 | 3 | 5 | 4 | 3 | 19 | 76 |
| 6 | The fonts used made the reading easier. | 4 | 4 | 5 | 4 | 4 | 21 | 84 |
| 7 | The colors used made reading easier. | 5 | 4 | 4 | 5 | 4 | 22 | 88 |
| 8 | I completed the tasks effectively. | 4 | 5 | 3 | 5 | 5 | 22 | 88 |
| 9 | The successive pages were easy to follow and understand. | 5 | 4 | 4 | 4 | 4 | 21 | 84 |
| 10 | I am satisfied with the website in terms of use. | 5 | 4 | 5 | 5 | 5 | 24 | 96 |

Admins were satisfied with the visual design of our website with a percentage of 76, which is relatively lower than other items. Admins are authorized to do many actions such as adding news and announcements to the website, doing the necessary updates, listening to the sound files submitted and approving membership applications. Therefore, it is projected that the number of the feedback given by the participants about the visual design of the website will increase in time due to further experiences in use. In addition, website design is based on the opinions of the visually impaired to ensure convenient and efficient use. Thus, the functionality was priotorized to visual elements for the visually-impaired.

**Table 7.** Interface for the Visually Impaired Satisfaction Survey Items and the data Obtained

|  | Items | A | B | C | D | E | TOTAL | % |
|---|---|---|---|---|---|---|---|---|
| 1 | I was easily be able to manage the website functions. | 4 | 4 | 5 | 4 | 4 | 21 | 84 |

| 2 | It is easy to learn how to use website effectively. | 4 | 4 | 4 | 3 | 5 | 20 | 80 |
| 3 | "Help" messages are sufficient. | 3 | 4 | 4 | 4 | 5 | 20 | 80 |
| 4 | I was able to correct the mistakes very easily. | 4 | 3 | 5 | 4 | 4 | 20 | 80 |
| 5 | I completed the tasks effectively. | 4 | 4 | 5 | 5 | 4 | 22 | 88 |
| 6 | The successive pages were easy to follow. | 4 | 5 | 4 | 4 | 5 | 22 | 88 |
| 7 | I am satisfied with the website in terms of use. | 5 | 5 | 4 | 5 | 5 | 24 | 96 |

According to Table 7, the visually-impaired participants agree with the statements "It is easy to learn how to use website effectively", "help" menu is sufficient" and "I was able to correct the mistakes very easily" with a percentage of 80. This agreement was higher than 80% for the other items. When positive feedback given about the automation system by the participants is considered, we can say that the system has successfully achieved its objectives in parallel with the demands and needs stated by the visually-impaired.

## 5. CONCLUSION

Designed and developed within the framework of this current study, web-based audio library automation system provides an effective, user-friendly and fast audio book / library solution for the visually-impaired. In this respect, thanks to this system, visually-impaired individuals who have no opportunity to go to a library are able to access audio books via the internet technology / internet access. Three types of user profile in the system enables the establishment of the dynamic structure of the system and a general approach to ensure the administration and sustainability of the system in an effective way.

The studies conducted regarding "evaluation" and "test processes" reveal positive results of the system in terms of target audience and study objectives. At this point, especially visually-impaired individuals provided positive feedback about the efficiency and functionality of the system. Therefore; the current study is believed to contribute a lot to the related literature and further research studies to be conducted in the field.

The positive feedback obtained in this study motivated the authors for further follow-up studies. It is projected to improve system features through different programming approaches in terms of specific item-based data obtained especially during evaluation – test processes. At this point, it is planned to improve the functionality and certain features of the system with the help of artificial intelligence-based smart system algorithms. Finally, the integration of the system into various languages and common use of the system in various institutional and organizational platforms as well as the evaluation process will be considered in the future studies as well.

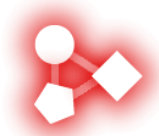